\documentstyle[11pt]{article}

\font\fr=eufm10 scaled \magstep 1 
\font\es=msbm11                  
\newtheorem{teor}{Theorem}

\newtheorem{definition}{Definition}
\def\beq{\begin{equation}}
\def\eeq{\end{equation}}
\def\bea{\begin{eqnarray}}
\def\eea{\end{eqnarray}}
\def\beann{\begin{eqnarray*}}
\def\eeann{\end{eqnarray*}}
\def\ben{\begin{enumerate}}
\def\een{\end{enumerate}}
\def\bit{\begin{itemize}}
\def\eit{\end{itemize}}
\def\dst{\(\displaystyle}
\def\derpar#1#2{\frac{\partial{#1}}{\partial{#2}}}

\def\feble#1{\mathrel{\mathop =\limits_{#1}}}

\def\moment#1#2#3{{#1}_{#2}, \ldots, {#1}_{#3}}

\def\qed{\ifvmode\removelastskip\fi
{\unskip\nobreak\hfil\penalty50\hbox{}\nobreak\hfil
\hbox{\vrule height1.2ex width1.2ex}\parfillskip=0pt
\finalhyphendemerits=0 \par\smallskip}}
\def\vf{\mbox{\fr X}}
\def\df{{\mit\Omega}}
\def\Lag{{\cal L}}

\def\d{{\rm d}}
\def\Real{\mbox{\es R}}

\def\inn{\mathop{i}\nolimits}
\def\Tan{{\rm T}}

\def\ls{(J^1E,\Omega_\Lag )}
\def\hs{(J^{1*}E,\Omega_h)}

\def\lag{\pounds}
\def\Cinfty{{\rm C}^\infty}
\def\proof{( {\sl Proof} )\quad}
\def\tabaddress#1{{\small\it\begin{tabular}[t]{c}#1 \\[1.2ex]\end{tabular}}}

\title{A GEOMETRICAL ANALYSIS OF THE FIELD EQUATIONS IN FIELD THEORY}
\author{\sc A. Echeverr\'{\i}a-Enr\'\i quez,
M.C. Mu\~noz-Lecanda
\thanks{{\bf e}-{\it mail}: matmcml@mat.upc.es},
N. Rom\'an-Roy
\thanks{{\bf e}-{\it mail}: matnrr@mat.upc.es}
   \\
  \tabaddress{\it Departamento de Matem\'atica Aplicada IV.\\
 Edificio C-3, Campus Norte UPC.\\
   C/ Jordi Girona 1.
   E-08034 Barcelona. SPAIN}}
\date{May 10, 2001}

\begin{document}

\maketitle

\begin{abstract}
In this review paper we give a geometric formulation
of the field equations in the Lagrangian and Hamiltonian formalisms
of classical field theories (of first order) in terms of multivector fields.
This formulation enables us to discuss the existence and non-uniqueness
of solutions of these equations, as well as their integrability.
\end{abstract}

\vspace{2cm}

\vspace{1cm}
\vbox{\raggedleft AMS s.\,c.\,(2000):
 53C05, 53C80, 55R10, 55R99, 58A20, 70S05.\\
PACS (1999): 02.40.Hw, 02.40.Vh, 11.10.Ef, 45.10.Na }\null

\clearpage

\section{Introduction}

In recent years new developments have been done in
the study of multisymplectic Hamiltonian systems
\cite{2} and in particular, it application
to describe field theories 
In this study, {\sl multivector fields} and their contraction with
differential forms are used; and this is an intrinsic formulation of the
systems of partial differential equations locally describing the field.
Thus, the integrability of such equations; that is, of multivector fields,
is a matter of interest. Given a fiber bundle $\pi\colon E\to M$,
certain integrable multivector fields in $E$
are equivalent to integrable connections in $E\to M$ \cite{6}.
 This result is applied in two particular situations:
\bit
\item
First considering multivector fields in $J^1E$
(the first-order jet bundle), in order to characterize
 integrable multivector fields
 whose integral manifolds are holonomic.
\item
 Second, considering the manifold
 $ J^{1*}E\equiv\Lambda_1^m\Tan^*E/\Lambda^m_0\Tan^*E$
 (where $\Lambda_1^m\Tan^*E$
 is the bundle of $m$-forms on
 $E$ vanishing by the action of two $\pi$-vertical vector fields,
 and $\Lambda^m_0\Tan^*E\equiv\pi^*\Lambda^m\Tan^*M$),
 wich is also a fiber bundle $J^{1*}E\to M$. Then,
 we will take multivector fields in $J^{1*}E$, in order to characterize
 those of them being integrable.
\eit
 From these results we can set the Lagrangian and Hamiltonian equations
for {\sl multisymplectic models} of first-order classical field theories
in a geometrical way \cite{3}, \cite{10}, \cite{12}, \cite{17},
in terms of multivector fields; which is equivalent to other formulations
using {\sl Ehresmann connections} in a jet bundle
\cite{15}, \cite{19}, or their associated
 {\sl jet fields} \cite{5}.
This formulation allows us to discuss several aspects of these equations,
in particular, the existence and non-uniqueness of solutions.
(See also \cite{13}, \cite{14}, where multivector fields are used
in a more specific context).

The structure of the work is the following:
In section 2 we introduce the terminology and nomenclature concerning with
multivector fields in differentiable manifolds and fiber bundles.
This is used in Section 3 for setting the field equations for
Lagrangian field theories (of first-order) in terms of multivector fields,
and for analizing their characteristic features.
Finally, the same study is made in Section 4 for Hamiltonian field theories.

 Thoughout this paper 
 $\pi\colon E\to M$ will be a fiber bundle ($\dim\, M=m$, $\dim\, E=N+m$),
 where $M$ is an oriented manifold with volume form $\omega\in\df^m(M)$.
 $\pi^1\colon J^1E\to E$ is the
 jet bundle of local sections of $\pi$, and
 $\bar\pi^1=\pi\circ\pi^1\colon J^1E \longrightarrow M$
 gives another fiber bundle structure.
 $(x^\mu,y^A,v^A_\mu)$ will denote natural local systems
 of coordinates in $J^1E$, adapted to the bundle $E\to M$
 ($\mu = 1,\ldots,m$; $A= 1,\ldots,N$), and such that
 $\omega=\d x^1\wedge\ldots\wedge\d x^m\equiv\d^mx$.

 Manifolds are real, paracompact,
 connected and $C^\infty$. Maps are $C^\infty$. Sum over crossed repeated
 indices is understood.

\section{Multivector fields in differentiable manifolds}
\protect\label{mvfdm}

Let $E$ be a $n$-dimensional differentiable manifold.
Sections of $\Lambda^m(\Tan E)$ are called
$m$-{\sl multivector fields} in $E$
(they are contravariant skewsymmetric tensors of order $m$ in $E$).
 Then, {\sl contraction} with multivector fields is the usual one for
 tensor fields in $J^{1*}E$.
We will denote by $\vf^m (E)$ the set of $m$-multivector fields in $E$.

If $Y\in\vf^m(E)$, for every $p\in E$, there exists an open neighbourhood
 $U_p\subset E$ and $Y_1,\ldots ,Y_r\in\vf (U_p)$ such that
\dst Y\feble{U_p}\sum_{1\leq i_1<\ldots <i_m\leq r}
f^{i_1\ldots i_m}Y_{i_1}\wedge\ldots\wedge Y_{i_m}\) ;
with $f^{i_1\ldots i_m}\in\Cinfty (U_p)$ and $m\leq r\leq{\rm dim}\, E$.
Then, $Y\in\vf^m(E)$ is said to be {\sl locally decomposable} if,
for every $p\in E$, there exists an open neighbourhood $U_p\subset E$
and $Y_1,\ldots ,Y_m\in\vf (U_p)$ such that
$Y\feble{U_p}Y_1\wedge\ldots\wedge Y_m$.

A non-vanishing $m$-multivector field $Y\in\vf^m(E)$ and
a $m$-dimensional distribution $D\subset\Tan E$
are {\sl locally associated} if there exists a connected open set
$U\subseteq E$ such that $Y\vert_U$ is a section of $\Lambda^mD\vert_U$.
If $Y,Y'\in\vf^m(E)$ are non-vanishing multivector fields
locally associated with the same distribution $D$,
on the same connected open set $U$, then there exists a
non-vanishing function $f\in\Cinfty (U)$ such that
$Y'\feble{U}fY$. This fact defines an equivalence relation in the
set of non-vanishing $m$-multivector fields in $E$, whose equivalence classes
will be denoted by $\{ Y\}_U$. Then:

\begin{teor}
There is a one-to-one correspondence between the set of $m$-dimensional
orientable distributions $D$ in $\Tan E$ and the set of the
equivalence classes $\{ Y\}_E$ of non-vanishing, locally decomposable
$m$-multivector fields in $E$.
\end{teor}
\proof
Let $\omega\in\df^m(E)$ be an orientation form for $D$.
If $p\in E$ there exists an open neighbourhood $U_p\subset E$
and $Y_1,\ldots ,Y_m\in\vf (U_p)$,
with $\inn(Y_1\wedge\ldots\wedge Y_m)\omega >0$, such that
\dst D\vert_{U_p}={\rm span}\, \{Y_1,\ldots ,Y_m\}\) .
Then $Y_1\wedge\ldots\wedge Y_m$ is a representative of a class
of $m$-multivector fields associated with $D$ in $U_p$.
But the family $\{ U_p\ ;\ p\in E\}$ is a covering of $E$;
let $\{ U_\alpha\ ;\ \alpha\in A\}$ be
a locally finite refinement and $\{ \rho_\alpha\ ;\ \alpha\in A\}$
a subordinate partition of unity.
If $Y^\alpha_1,\ldots ,Y^\alpha_m$ is a local basis of $D$ in $U_\alpha$,
with $\inn(Y^\alpha_1\wedge\ldots\wedge Y^\alpha_m)\omega >0$,
then \dst Y=\sum_\alpha\rho_\alpha Y^\alpha_1\wedge\ldots\wedge Y^\alpha_m\)
is a global representative of the class of non-vanishing
$m$-multivector fields associated with $D$ in $E$.

The converse is trivial because, if
$Y\vert_U=Y^1_1\wedge\ldots\wedge Y^1_m=Y^2_1\wedge\ldots\wedge Y^2_m$,
for different sets $\{ Y^1_1,\ldots ,Y^1_m\}$, $\{Y^2_1,\ldots ,Y^2_m\}$,
then
$span\, \{ Y^1_1,\ldots ,Y^1_m\} =span\, \{ Y^2_1,\ldots ,Y^2_m\}$.
\qed

If $Y\in\vf^m(E)$ is non-vanishing and locally decomposable; and
$U\subseteq E$ is a connected open set, the distribution associated
with the class $\{ Y\}_U$ is denoted ${\cal D}_U(Y)$.
If $U=E$ we write ${\cal D}(Y)$.

A non-vanishing, locally decomposable
multivector field $Y\in\vf^m(E)$ is said to be {\sl integrable}
(resp. {\sl involutive}) if 
 it associated distribution ${\cal D}_U(Y)$ is integrable
(resp. involutive).
Of course, if $Y\in\vf^m(E)$ is integrable (resp. involutive),
then so is every other in it equivalence class $\{ Y\}$,
and all of them have the same integral manifolds.
Moreover, the {\sl Frobenius' theorem} allows us to say that
a non-vanishing and locally decomposable multivector field is integrable 
 if, and only if, it is involutive. 
Nevertheless, in many applications, we have locally decomposable
multivector fields $Y\in\vf^m(E)$ which are not integrable in $E$;
 but integrable in a submanifold of $E$.
A (local) algorithm for finding this submanifold
 has been developed \cite{6}.

The particular situation which we will pay attention
is the study of multivector fields in fiber bundles.
Then, if $\pi\colon E\to M$ is a fiber bundle ,
we will be interested in the case that the integral manifolds of
integrable multivector fields in $E$ are sections of $\pi$.
Thus, $Y\in\vf^m(E)$ is said to be {\sl $\pi$-transverse}
if, at every point $y\in E$,
$(\inn (Y)(\pi^*\omega))_y\not= 0$, for every $\omega\in\df^m(M)$
with $\omega (\pi(y))\not= 0$.
Then, if $Y\in\vf^m(E)$ is integrable, it is $\pi$-transverse if, and only if,
it integral manifolds are local sections of $\pi\colon E\to M$.
In this case, if $\phi\colon U\subset M\to E$
is a local section with $\phi (x)=y$ and $\phi (U)$ is
the integral manifold of $Y$ through $y$,
then $\Tan_y({\rm Im}\,\phi)$ is ${\cal D}_y(Y)$.

\section{Lagrangian equations in classical field theories}

A {\sl classical field theory} is described by it {\sl configuration bundle}
$\pi\colon E\to M$;
and a {\sl Lagrangian density} which is
a $\bar\pi^1$-semibasic $m$-form on $J^1E$.
A Lagrangian density is usually written as
$\Lag =\lag (\bar\pi^{1^*}\omega)$, where $\lag\in\Cinfty (J^1E)$
is the {\sl Lagrangian function} associated with $\Lag$ and $\omega$.

The {\sl Poincar\'e-Cartan $m$ and $(m+1)$-forms}
associated with the Lagrangian density $\Lag$
are defined using the {\sl vertical endomorphism}
 ${\cal V}$ of the bundle $J^1E$:
$$
\Theta_{\Lag}:=\inn({\cal V})\Lag+\Lag\in\df^{m}(J^1E)
\quad ;\quad
\Omega_{\Lag}:= -\d\Theta_{\Lag}\in\df^{m+1}(J^1E)
$$
Then a {\sl Lagrangian system} is a couple $\ls$.
The Lagrangian system is {\sl regular} if
 $\Omega_{\Lag}$ is $1$-nondegenerate.
In a natural chart in $J^1E$ we have
\bea
\Omega_{\Lag}=
-\frac{\partial^2\lag}{\partial v^B_\nu\partial v^A_\mu}
\d v^B_\nu\wedge\d y^A\wedge\d^{m-1}x_\mu 
-\frac{\partial^2\lag}{\partial y^B\partial v^A_\mu}\d y^B\wedge
\d y^A\wedge\d^{m-1}x_\mu +
\nonumber \\ 
\frac{\partial^2\lag}{\partial v^B_\nu\partial v^A_\mu}v^A_\mu
\d v^B_\nu\wedge\d^mx  +
\left(\frac{\partial^2\lag}{\partial y^B\partial v^A_\mu}v^A_\mu
 -\derpar{\lag}{y^B}+
\frac{\partial^2\lag}{\partial x^\mu\partial v^B_\mu}
\right)\d y^B\wedge\d^mx
\label{omegalag}
\eea
(where \dst\d^{m-1}x_\mu\equiv\inn\left(\derpar{}{x^\mu}\right)\d^mx\) );
and the regularity condition is equivalent to
\dst det\left(\frac{\partial^2\lag}
{\partial v^A_\mu\partial v^B_\nu}(\bar y)\right)\not= 0\) ,
for every $\bar y\in J^1E$.

A variational problem can be stated for $\ls$ ({\sl Hamilton principle}\/):
the states of the field are the sections of $\pi$ (denoted by $\Gamma(M,E)$)
 which are critical for the functional
${\bf L}\colon\Gamma(M,E)\to\Real$ defined by
${\bf L}(\phi):=\int_M(j^1\phi)^*\Lag$, for every $\phi\in\Gamma(M,E)$.
These critical sections can be characterized by the condition
 $$
 (j^1\phi)^*\inn (X)\Omega_\Lag=0 \quad ,
 \mbox{\rm for every $X\in\vf (J^1E)$}
 $$
 In natural coordinates, if $\phi=(x^\mu,y^A(x))$,
 this condition is equivalent to demanding that
 the components o
f $\phi$ satisfy the {\sl Euler-Lagrange equations}
 \beq
 \derpar{\lag}{y^A}\Big\vert_{j^1\phi}-
\derpar{}{x^\mu}\derpar{\lag}{v_\mu^A}\Big\vert_{j^1\phi} = 0
 \quad , \quad \mbox{\rm (for $A=1,\ldots ,N$)}
 \label{eleq}
 \eeq
(For a more detailed description on all these concepts see, for instance,
 \cite{1}, \cite{3}, \cite{5}, \cite{9}, \cite{10},
 \cite{11}, \cite{18}, \cite{19}).

 The problem of finding these critical sections
 can be formulated equivalently as follows:
 to finding a distribution $D$ of $\Tan (J^1E)$ satisfying that:
\bit
\item
 $D$ is integrable (that is, {\sl involutive\/}).
\item
 $D$ is $m$-dimensional.
\item
 $D$ is $\bar\pi^1$-transverse.
\item
 The integral manifolds of $D$ are the critical sections of the Hamilton
 principle.
\eit

 Then, from the first and second conditions, there exist
 $\moment{X}{1}{m}\in\vf(J^1E)$ (in involution),
 which locally span $D$. Therefore
 $X=X_1\wedge\ldots\wedge X_m$ defines a section of
 $\Lambda^m\Tan (J^1E)$, that is, a non-vanishing, locally decomposable
 multivector field in $J^1E$, whose local expression in natural coordinates is
 \beq
 X=\bigwedge_{\mu=1}^m f_\mu\left(\derpar{}{x^\mu}+F_\mu^A\derpar{}{y^A}+
 G_{\mu\rho}^A\derpar{}{v_\rho^A}\right)
 \label{locmvf}
 \eeq
 where $f_\mu$ are non-vanishing functions.
 A representative of the class $\{ X\}$ can be selected by the condition
 $\inn (X)(\bar\pi^{1*}\omega)=1$ which, as a particular solution,
 leads to $f_\mu=1$, for every $\mu$.
 Furthermore, the third and fourth conditions impose that $X$ is
 $\bar\pi^1$-transverse, integrable and
 it integral manifolds are holonomic sections of $\bar\pi^1$.

 Bearing this in mind, we want to characterize the
 integrable multivector fields in $J^1E$
 whose integral manifolds are canonical prolongations of sections of $\pi$. 
 So, consider the vector bundle projection
 $\kappa\colon\Tan J^1E\to\Tan E$ defined by
 $$
 \kappa (\bar y,\bar u):=
 \Tan_{\bar\pi^1(\bar y)}\phi (\Tan_{\bar y}\bar\pi^1(\bar u))
 \qquad
 \mbox{\rm where $(\bar y,\bar u)\in\Tan J^1E$, $\phi\in\bar y$}
 $$
 This projection is extended in a natural way to
 $\Lambda^m\kappa\colon\Lambda^m\Tan J^1E\to\Lambda^m\Tan E$.
 Then, a $\bar\pi^1$-transverse multivector field $X\in\vf^m(J^1E)$
 is said to be {\sl semi-holonomic}, or a
 {\sl Second Order Partial Differential Equation},
 if $\Lambda^m\kappa\circ X=\Lambda^m\Tan\pi^1\circ X$.
 In a natural chart in $J^1E$, the local expression of $X$ is
 $$
 X\equiv
 \bigwedge_{\mu=1}^m f_\mu\left(\derpar{}{x^\mu}+v_\mu^A\derpar{}{y^A}+
 G_{\mu\rho}^A\derpar{}{v_\rho^A}\right)
 $$
 where $f_\mu\in\Cinfty(J^1E)$ are arbitrary non-vanishing functions.
 On the other hand, $X\in\vf^m(J^1E)$ is said to be {\sl holonomic} if
 it is integrable, $\bar\pi^1$-transverse and
 it integral sections $\psi\colon M\to J^1E$ are holonomic.
 Then, it can be proved \cite{6}
 that a multivector field
 $X\in\vf^m(J^1E)$ is holonomic if, and only if,
 it is integrable and semi-holonomic.

 Of course, if $X\in\vf^m(J^1E)$ is a semi-holonomic
 (resp. holonomic) multivector field, everyone in the class
 $\{ X\}\subset\vf^m(J^1E)$ are semi-holonomic (resp. holonomic) too.
 As local expression of a representative we can take
 \beq
 X\equiv
 \bigwedge_{\mu=1}^m \left(\derpar{}{x^\mu}+v_\mu^A\derpar{}{y^A}+
 G_{\mu\rho}^A\derpar{}{v_\rho^A}\right)
 \label{locsopde}
 \eeq
 Then, given a section $\phi =( x^\mu ,f^A)$,
 if \dst j^1\phi =\left( x^\mu ,f^A,\derpar{f^A}{x^\rho}\right)\)
 is an integral section of this semi-holonomic multivector field,
 then \dst v^A_\mu=\derpar{f^A}{x^\mu}\) , and the components of
 $\phi$ are solution of the system of partial differential equations
 \beq
 G_{\nu\rho}^A\left(x^\mu ,f^A,\derpar{f^A}{x^\mu}\right) =
 \frac{\partial^2f^A}{\partial x^\rho\partial x^\nu}
 \label{insec}
 \eeq
 On the other hand, it can be proved
 \cite{6} that
 classes of locally decomposable and $\bar\pi^1$-transverse
 multivector fields are in one-to-one correspondence
 with orientable connections in the bundle
 $\pi\colon J^1E\to M$ (this correspondence is characterized by the fact that
 ${\cal D}(X)$ is the {\sl horizontal subbundle} of the connection).
 For the multivector field (\ref{locsopde}), the associated
 Ehresmann connection has the local expression
 $$
 \nabla=\d x^\mu\otimes
 \left(\derpar{}{x^\mu}+v_\mu^A\derpar{}{y^A}+
 G_{\mu\rho}^A\derpar{}{v_\rho^A}\right)
 $$
 Then $X\in\vf^m(J^1E)$ is integrable if, and only if,
 the connection $\nabla$ associated with the class $\{ X\}$ is {\sl flat};
 that is, the curvature of $\nabla$ vanishes everywhere. Thus,
 the system (\ref{insec}) has solution
 if, and only if, the following additional system of equations holds
(for every $B,\mu,\rho,\eta$)
\beq
\left.
\begin{array}{lll}
0 &=& G^B_{\eta\mu}-G^B_{\mu\eta}
\\
0 &=& \derpar{G_{\eta\rho}^B}{x^\mu}+v_\mu^A\derpar{G_{\eta\rho}^B}{y^A}+
G_{\mu\gamma}^A\derpar{G_{\eta\rho}^B}{v_\gamma^A}-
\derpar{G_{\mu\rho}^B}{x^\eta}-v_\eta^A\derpar{G_{\mu\rho}^B}{y^A}-
G_{\eta\gamma}^A\derpar{G_{\mu\rho}^B}{v_\gamma^A}
\end{array} \right\}
\label{curvcero}
\eeq

 Now, the problem posed by the Hamilton principle
 can be stated in the following way:

\begin{teor}
Let $\ls$ be a Lagrangian system.
The critical sections of the Lagrangian variational problem
are the integral sections of a class of
holonomic multivector fields $\{ X_{\Lag}\}\subset\vf^m(J^1E)$, such that
 $$
 \inn (X_{\Lag})\Omega_{\Lag}=0
 \quad , \quad
 \mbox{\rm for every $X_\Lag\in\{ X_{\Lag}\}$}
 $$
\label{important}
\end{teor}
\proof
The critical sections must be the integral sections of a class of
holonomic multivector fields $\{ X_{\Lag}\}\subset\vf^m(J^1E)$,
as a consequence of the above discussion.

Now, using the local expression (\ref{omegalag}) of $\Omega_\Lag$,
and taking (\ref{locsopde}) as the
representative of the class of semi-holonomic multivector fields
$\{ X_{\Lag}\}$, from the relation
$\inn (X_{\Lag})\Omega_{\Lag}=0$ we have that the
coefficients on $\d v^A_\mu$, $\d y^A$ and $\d x^\mu$ must vanish.
But, for the coefficients on $\d v^A_\mu$ we obtain the identities
$$
0 = (v^B_\mu-v^B_\mu)\frac{\partial^2\lag}{\partial v^A_\nu\partial v^B_\mu}
\qquad (\mbox{for every $A,\nu$})
$$
meanwhile the condition for the coefficients on $\d y^A$
 leads to the system of equations
\beq
\frac{\partial^2\lag}{\partial v^B_\nu\partial  v^A_\mu}G^B_{\nu\mu}=
\derpar{\lag}{y^A}-\frac{\partial^2\lag}{\partial x^\mu\partial  v^A_\mu}-
\frac{\partial^2\lag}{\partial y^B\partial v^A_\mu}v^B_\mu
\label{eqsG}
\qquad (A=1,\ldots ,N)
\eeq
Therefore, if \dst j^1\phi=\left( x^\mu,f^A,\derpar{f^A}{x^\nu}\right)\)
must be an integral section of $X_\Lag$, then
\dst v^A_\mu=\derpar{f^A}{x^\mu}\) , and hence
the coefficients $G^B_{\nu\mu}$ must satisfy equations (\ref{insec}) .
As a consequence, the system (\ref{eqsG}) is equivalent 
to the Euler-Lagrange equations for the section $\phi$.
Note that, from the above conditions, the coefficients on
 $\d x^\mu$ vanish identically.
\qed

So, in Lagrangian field theories,
we search for (classes of) non-vanishing and locally decomposable
 multivector fields $X_{\Lag}\in\vf^m(J^1E)$ such that:
\ben
\item
The equation $\inn (X_{\Lag})\Omega_{\Lag}=0$ holds.
\item
$X_{\Lag}$ are semi-holonomic.
\item
$X_{\Lag}$ are integrable.
\een
Then we introduce the following nomenclature:

\begin{definition}
 $X_{\Lag}\in\vf^m(J^1E)$ is said to be an
{\rm Euler-Lagrange multivector field} for $\Lag$ if it is semi-holonomic and
is a solution of the  equation $\inn (X_{\Lag})\Omega_{\Lag}=0$.
\end{definition}

Observe that neither the compatibility of the system
(\ref{eqsG}), nor the integrability of (\ref{insec}) are assured.
Thus, the existence of Euler-Lagrange multivector fields is not guaranteed
in general, and, if they exist, they are not integrable necessarily. Then:

\begin{teor}
{\rm (Existence and local multiplicity of Euler-Lagrange
 multivector fields)}.
Let $\ls$ be a regular Lagrangian system. Then:
\ben
\item
There exist classes of Euler-Lagrange multivector fields for $\Lag$.
\item
In a local system these multivector fields depend on
 $N(m^2-1)$ arbitrary functions.
\een
\label{holsecreg}
\end{teor}
\proof
\ben
\item
First we analyze the local existence of solutions
 and then their global extension.

In a chart of natural coordinates in $J^1E$,
using the local expression (\ref{omegalag}) of $\Omega_{\Lag}$,
and taking the multivector field given in (\ref{locmvf}) (with
$f_\mu=1$, for every $\mu$) as the representative of the class
 $\{ X_{\Lag}\}$, from the relation
$\inn (X_{\Lag})\Omega_{\Lag}=0$ we have that the
coefficients on $\d v^A_\mu$, $\d y^A$ and $\d x^\mu$ must vanish.

Thus, for the coefficients on $\d v^A_\mu$, we obtain that
$$
0 = (F^B_\mu-v^B_\mu)\frac{\partial^2\lag}{\partial v^A_\nu\partial v^B_\mu}
\qquad (\mbox{for every $A,\nu$})
$$
But, if $\Lag$ is regular, the matrix
\dst\left(\frac{\partial^2\lag}{\partial v^A_\nu\partial v^B_\mu}\right)\)
is regular. Therefore $F^B_\mu=v^B_\mu$ (for every $B,\mu$);
which proves that if $X_{\Lag}$ exists it is semi-holonomic.

Afterwards, from the condition for the coefficients on
$\d y^A$, and taking into account that we have obtained $F^B_\mu=v^B_\mu$,
 we obtain the equations (\ref{eqsG}),
which is a system of $N$ linear equations on the functions $G^B_{\nu\mu}$.
This is a compatible system as a consequence of the regularity of $\Lag$,
since the matrix of the coefficients has (constant) rank equal to $N$
(observe that the matrix of this system is obtained as a rearrangement
of rows of the Hessian matrix).

From the above conditions, we obtain that the coefficients on
$\d x^\mu$ vanish identically.

These results allow us to assure the local existence of
 (classes of) multivector fields
satisfying the desired conditions. The corresponding global solutions
are then obtained using a partition of unity subordinated
to a covering of $J^1E$ made of local natural charts.
\item
The expression of a semi-holonomic multivector field
$X_{\Lag}\in\{ X_{\Lag}\}$ is given by (\ref{locsopde}).
So, it is determined by the $Nm^2$ coefficients $G^B_{\nu\mu}$,
which are related  by the $N$ independent equations (\ref{eqsG}).
Therefore, there are $N(m^2-1)$ arbitrary functions.
\een
\qed

Now the problem is to finding a class of
 integrable Euler-Lagrange multivector field, if it exists.
So, we can choose from the solutions of this system, those such that
$X_{\Lag}$ verify the integrability condition;
that is, the associated connection $\nabla_\Lag$ is flat
 (equations (\ref{curvcero})).
If the equations (\ref{eqsG})
and the first group of equations (\ref{curvcero}) allow us to isolate
$N+\frac{1}{2}Nm(m-1)$ coefficients $G^A_{\mu\nu}$
 as functions on the remaining ones;
and the set of $\frac{1}{2}Nm^2(m-1)$ partial differential equations
(the second group of equations (\ref{curvcero}))
 on these remaining coefficients
satisfies the conditions on
 {\sl Cauchy-Kowalewska's theorem} \cite{4},
then the existence of integrable Euler-Lagrange multivector fields is assured.

 \bit
 \item
  {\bf Remark}: ({\sl Singular Lagrangian systems\/})

For singular Lagrangian systems,
the existence of Euler-Lagrange multivector fields
 is not assured except perhaps
on some submanifold $S\hookrightarrow J^1E$.
Even more, locally decomposable and $\bar\pi^1$-transverse
multivector fields, solutions of the field equations
can exist (in general, on some submanifold of $J^1E$),
but none of them being semi-holonomic (at any point of this submanifold).
As in the regular case, although Euler-Lagrange multivector fields exist
on some submanifold $S$, their integrability is not assured except
perhaps on another smaller submanifold $I\hookrightarrow S$; such that
the integral sections are contained in $I$.
This condition implies that $\bar\pi^1\vert_I\colon I\to M$
 must be onto on $M$.

The local treatment of the singular case is as follows:
starting from (\ref{locmvf}), and taking the representative obtained
 by making $f_\mu=1$, for every $\mu$,
we can impose the semi-holonomic condition by making $F^A_\mu=v^A_\mu$,
for every $A,\mu$. Therefore, we have the system of equations (\ref{eqsG})
for the coefficients $G^A_{\mu\nu}$;
but this system is not compatible in general
except perhaps in a set of points $S_1\subset J^1E$,
which is assumed to be a non-empty closed submanifold.
Then, there are Euler-Lagrange multivector fields on $S_1$,
but the number of arbitrary functions on which they depend
is not the same as in the regular case, since it depends on the dimension of
$S_1$ and the rank of the Hessian matrix of $\lag$.
Next, the tangency condition must be analyze; and
finally the question of integrability must be considered as above,
but for a submanifold of $S_1$.
\eit

\section{Hamiltonian equations in classical field theories}

For the Hamiltonian formalism of field theories,
the choice of a {\sl multimomentum phase space}
 or {\sl multimomentum bundle} is not unique
 (see \cite{8}). In this work we take:
 $$
 J^{1*}E\equiv\Lambda_1^m\Tan^*E/\Lambda^m_0\Tan^*E
 $$
 (where $\Lambda_1^m\Tan^*E$
 is the bundle of $m$-forms on
 $E$ vanishing by the action of two $\pi$-vertical vector fields,
 and $\Lambda^m_0\Tan^*E\equiv\pi^*\Lambda^m\Tan^*M$).
 We have the natural projections
 $$
 \tau^1\colon J^{1*}E\to E \quad ,\quad
 \bar\tau^1=\pi\circ\tau^1\colon J^{1*}E\to M
 $$
 and we denote by  $(x^\mu,y^A,p_A^\mu)$ the
 natural local systems of coordinates in $J^{1*}E$ adapted to these bundle
 structures ($\mu = 1,\ldots,m$; $A= 1,\ldots,N$).

 For constructing  {\sl Hamiltonian systems},
 $J^{1*}E$  must be endowed with a geometric structure.
 There are different ways for making this, namely:
 using {\sl Hamiltonian sections}, or {\sl Hamiltonian densities}
 \cite{3}, \cite{8}, \cite{10}.
 So we construct the {\sl Hamilton-Cartan} $m$ and $(m+1)$ {\sl forms}
 $\Theta_h\in\df^m(J^{1*}E)$, and
 $\Omega_h=-\d\Theta_h\in\df^{m+1}(J^{1*}E)$,
 which have the local expressions (in an open set $U\subset J^{1*}E$)
 \bea
 \Theta_h &=& p_A^\mu\d y^A\wedge\d^{m-1}x_\mu -H\d^mx
 \nonumber \\
 \Omega_h &=& -\d p_A^\mu\wedge\d y^A\wedge\d^{m-1}x_\mu +
 \d H\wedge\d^mx
 \label{omegaH}
 \eea
 where $H\in\Cinfty (U)$ is a {\sl local Hamiltonian function}.
 A couple $\hs$ is said to be a {\sl Hamiltonian system}.

 We can state a variational problem for $(J^{1*}E,\Omega_h)$
 ({\sl Hamilton-Jacobi principle}\/):
 the states of the field are the sections of
 $\bar\tau^1$ which are critical for the functional
 \dst{\bf H}(\psi):=\int_M\psi^*\Theta_h\) ,
 for every $\psi\in\Gamma(M,J^{1*}E)$.
 They are characterized by the condition
 \cite{3}, \cite{8}
 $$
 \psi^*\inn (X)\Omega_h=0 \quad ,
 \mbox{\rm  for every $X\in\vf (J^{1*}E)$}
 $$
 In natural coordinates, if
 $\psi(x)=(x^\mu,y^A(x),p^\mu_A(x))$, this condition leads to the system
 \beq
 \derpar{y^A}{x^\mu}\Bigg\vert_{\psi}=
 \derpar{H}{p^\mu_A}\Bigg\vert_{\psi}
 \quad ;\quad
 \derpar{p_A^\mu}{x^\mu}\Bigg\vert_{\psi}=
 -\derpar{H}{y^A}\Bigg\vert_{\psi}
 \label{HDWeqs}
\eeq
 which is known as the {\sl Hamilton-De Donder-Weyl equations}.

 Let $\hs$ be a Hamiltonian system.
 The problem of finding critical sections solutions of the
 Hamilton-Jacobi principle can be formulated equivalently as follows:
 to finding a distribution $D$ of $\Tan (J^{1*}E)$
 satisfying that:
\bit
\item
 $D$ is integrable (that is, {\sl involutive\/}).
\item
 $D$ is $m$-dimensional.
\item
 $D$ is $\bar\tau^1$-transverse.
\item
 The integral manifolds of $D$ are the critical sections of
 the Hamilton-Jacobi principle.
\eit

 Then, from the first and the second conditions, there exist
 $\moment{X}{1}{m}\in\vf(J^{1*}E)$ (in involution),
 which locally span $D$. Therefore
 $X=X_1\wedge\ldots\wedge X_m$ defines a section of
 $\Lambda^m\Tan (J^{1*}E)$, that is, a non-vanishing, locally decomposable
 multivector field in $J^{1*}E$,
 whose local expression in natural coordinates is
 \beq
 X=\bigwedge_{\mu=1}^m
 f_\mu\left(\derpar{}{x^\mu}+F_\mu^A\derpar{}{y^A}+
 G^\rho_{A\mu}\derpar{}{p^\rho_A}\right)
 \label{locmvf2}
 \eeq
  where $f_\mu\in\Cinfty(J^{1*}E)$ are non-vanishing functions.
 A representative of the class $\{ X\}$ can be selected by the condition
 $\inn (X)(\bar\tau^{1*}\omega)=1$ which, as a particular solution,
 leads to $f_\mu=1$, for every $\mu$.

Therefore, the problem posed by the Hamilton-Jacobi principle
 can be stated in the following way:

 \begin{teor}
 The critical sections of the Hamilton-Jacobi principle are
 the sections $\psi\in\Gamma_c(M,J^{1*}E)$ such that
 they are the integral sections of a class of integrable and
 $\bar\tau^1$-transverse multivector fields
 $\{ X_{\cal H}\}\subset\vf^m(J^{1*}E)$ satisfying that
 $$
 \inn (X_{\cal H})\Omega_h=0 \quad ,   \quad
 \mbox{\rm for every $X_{\cal H}\in\{ X_{\cal H}\}$}
 $$
 \label{hameq}
 \end{teor}
\proof
The critical sections must be the integral sections of a class of
integrable and $\bar\tau^1$-transverse multivector fields
 $\{ X_{\cal H}\}\subset\vf^m(J^{1*}E)$,
as a consequence of the above discussion.

 Now, using the local expression (\ref{omegaH}) of
 $\Omega_h$; and taking the multivector field
 (\ref{locmvf2}) (with $f_\mu=1$, for every $\mu$)
 as a representative of the class $\{X_{\cal H}\}$,
 from $\inn (X_{\cal H})\Omega_h=0$ we obtain that
 the coefficients on $\d p_A^\mu$ must vanish:
 \beq
 0=F^A_\nu -\derpar{H}{p_A^\nu}
 \qquad (\mbox{for every $A,\nu$})
 \label{eqsG1}
 \eeq
 and the same happens for the coefficients on $\d y^A$:
 \beq
 0=G^\mu_{A\mu}+\derpar{H}{y^A}
 \qquad (A=1,\ldots ,N)
 \label{eqsG2}
 \eeq
 (Using these results, the coefficients on
 $\d x^\mu$ vanish identically).

 Now, if $\psi(x) =(x^\mu ,y^A(x^\nu ),p^\mu_A(x^\nu ))$
 has to be an integral section of $X_{\cal H}$ then
 $$
 F^A_\mu\circ \psi =\derpar{y^A}{x_\mu} \quad ; \quad
 G^\mu_{A\mu}\circ \psi = -\derpar{p^\mu_A}{x^\mu}
 $$
 and equations (\ref{eqsG1}) and
 (\ref{eqsG2}) are the Hamilton-De Donder-Weyl
 equations (\ref{HDWeqs}) for $\psi$.
 \qed

 Thus, we search for (classes of)
 $\bar\tau^1$-transverse and locally decomposable multivector fields
 $X_{\cal H}\in\vf^m(J^{1*}E)$ such that:
 \ben
 \item
 $\inn (X_{\cal H})\Omega_h=0$ holds.
 \item
 $X_{\cal H}$ are integrable.
 \een

 Classes of locally decomposable and $\bar\tau^1$-transverse
 multivector fields are in one-to one correspondence
 with connections in the bundle
 $\bar\tau^1\colon J^{1*}E\to M$.
 Then $X_{\cal H}$ is integrable if, and only if,
 the curvature of the connection associated
 with this class vanishes everywhere.

 \begin{definition}
 $X_{\cal H}\in\vf^m(J^{1*}E)$ will be called a
 {\rm Hamilton-De Donder-Weyl (HDW) multivector field}
 for the system $\hs$ if it is
 $\bar\tau^1$-transverse, locally decomposable and verifies the
 equation $\inn (X_{\cal H})\Omega_h=0$.
 \end{definition}

 For a Hamiltonian system, the existence of
 Hamilton-De Donder Weyl multivector fields is guaranteed,
 although they are not integrable necessarily.
 In fact:

 \begin{teor}
 {\rm (Existence and local multiplicity of HDW-multivector fields):}
 Let $\hs$ be a Hamiltonian system. Then
 \ben
 \item
 There exist classes of HDW-multivector fields $\{ X_{\cal H}\}$.
 \item
 In a local system the above solutions depend on $N(m^2-1)$
 arbitrary functions.
 \een
 \label{holsecreg2}
 \end{teor}
 \proof
 \ben
 \item
 Bearing in mind the proof of Theorem \ref{hameq},
 we have that the equations (\ref{eqsG1}) make a system of $Nm$
 linear equations which determines univocally the functions $F^A_\nu$,
 meanwhile the equations (\ref{eqsG2}) are
 a compatible system of $N$ linear equations on the
 $Nm^2$ functions $G^\mu_{A\nu}$.
 These results assure the local existence. The global solutions are
 obtained using a partition of unity subordinated to a
 covering of $J^{1*}E$ made of natural charts.
\item
 In natural coordinates in $J^{1*}E$,
 a representative of a class of HDW-multivector fields
 $X_{\cal H}\in\{ X_{\cal H}\}$ is
 given by (\ref{locmvf2}) (with $f_\mu=1$, for every $\mu$).
 So, it is determined by the $Nm$
 coefficients $F^A_\nu$, which are obtained as the solution of
 (\ref{eqsG1}), and by the $Nm^2$ coefficients
 $G^\mu_{A\nu}$, which are related  by the $N$ independent
 equations (\ref{eqsG2}). Therefore, there are $N(m^2-1)$ arbitrary
 functions.
 \een
 \qed

 For finding a class of integrable
 HDW-multivector fields (if it exists) we must impose
 that $X_{\cal H}$ verify the
 integrability condition: the curvature of the associated
 connection $\nabla_{\cal H}$ vanishes everywhere, that is,
 the following system of equations holds (for $1\leq\mu <\eta\leq m$)
 \bea 0 &=&
 \derpar{F_\eta^B}{x^\mu}+F_\mu^A\derpar{F_\eta^B}{y^A}+
 G^\gamma_{A\mu}\derpar{F_ \eta^B}{p^\gamma_A}-
 \derpar{F_\mu^B}{x^\eta}-F_\eta^A\derpar{F_\mu^B}{y^A}-
 G^\rho_{A\eta}\derpar{F_\mu^B}{p^\rho_A} \nonumber
 \\ & =&
 \frac{\partial^2H}{\partial x^\mu\partial p_B^\eta}+
 \derpar{H}{p_A^\mu}\frac{\partial^2 H}{\partial y^A\partial p_B^\eta}+
 \nonumber \\ & &
 G^\gamma_{A\mu}\frac{\partial^2H}{\partial
 p_A^\gamma\partial p_B^\eta}-
 \frac{\partial^2 H}{\partial x^\eta\partial p_B^\eta}-
 \derpar{H}{p_A^\eta}\frac{\partial^2H}{\partial y^A\partial
 p_B^\mu}- G^\rho_{A\eta}\frac{\partial^2H}{\partial
 p_A^\rho\partial p_B^\mu} \label{curvcero1}
 \\
 0 &=&
 \derpar{G^\rho_{B\eta}}{x^\mu}+F_\mu^A\derpar{G^\rho_{B\eta}}{y^A}+
 G^\gamma_{A\mu}\derpar{G^\rho_{B\eta}}{p^\gamma_A}-
 \derpar{G^\rho_{B\mu}}{x^\eta}-F_\eta^A\derpar{G^\rho_{B\mu}}{y^A}-
 G^\gamma_{A\eta}\derpar{G^\rho_{B\mu}}{p^\gamma_A} \nonumber
 \\ &=&
 \derpar{G^\rho_{B\eta}}{x^\mu}+
 \derpar{H}{p^\mu_A}\derpar{G^\rho_{B\eta}}{y^A}+
 \nonumber
 \\ & &
 G^\gamma_{A\mu}\derpar{G^\rho_{B\eta}}{p^\gamma_A}-
 \derpar{G^\rho_{B\mu}}{x^\eta}-
 \derpar{H}{p^\eta_A}\derpar{G^\rho_{B\mu}}{y^A}-
 G^\gamma_{A\eta}\derpar{G^\rho_{B\mu}}{p^\gamma_A}
 \label{curvcero2}
 \eea
 (where use is made of the Hamiltonian equations).
 Hence the number of arbitrary functions will
 be in general less than $N(m^2-1)$.

 As this is a system of partial differential
 equations with linear restrictions, there is no way of assuring
 the existence of an integrable solution.
 Considering the Hamiltonian equations  (\ref{eqsG2}) for the
 coefficients $G^\mu_{A\nu}$, together
 with the integrability conditions (\ref{curvcero1}) and
 (\ref{curvcero2}), we have \dst N+\frac{1}{2}Nm(m-1)\) linear
 equations and \dst\frac{1}{2}Nm^2(m-1)\) partial differential
 equations. Then, if the set of linear restrictions (\ref{eqsG2})
 and (\ref{curvcero1}) allow us to isolate \dst
 N+\frac{1}{2}Nm(m-1)\) coefficients $G^\mu_{A\nu}$ as functions on
 the remaining ones; and the set of \dst\frac{1}{2}Nm^2(m-1)\)
 partial differential equations (\ref{curvcero2}) on these
 remaining coefficients satisfies certain conditions,
 then the existence of
 integrable HDW-multivector fields (in $J^{1*}E$) is assured.
 If this is not the case, we can eventually select some particular
 HDW-multivector field solution, and apply an integrability
 algorithm in order to find a
 submanifold ${\cal I}\hookrightarrow J^{1*}E$ (if it exists),
 where this multivector field is integrable
 (and tangent to ${\cal I}$).

 {\bf Remarks}:
 \bit
 \item
 ({\sl Restricted Hamiltonian systems\/})

 There are many interesting cases in field theories
 where the Hamiltonian field equations are established
 not in $J^{1*}E$, but rather in a submanifold
 ${\rm j}_0\colon P\hookrightarrow J^{1*}E$,
 such that $P$ is a fiber bundle over $E$ (and $M$), and
 the corresponding projections
 $\tau^1_0\colon P\to E$ and $\bar\tau^1_0\colon P\to M$
 satisfy that $\tau^1\circ{\rm j}_0=\tau^1_0$ and
 $\bar\tau^1\circ{\rm j}_0=\bar\tau^1_0$.

 Now, not even the existence of
 HDW-multivector fields is assured, and
 an algorithmic procedure in order to obtain a submanifold
 $S_f$ of $P$
 where HDW-multivector fields exist, can be outlined.
 Of course the solution is not unique, in general,
 but the number of arbitrary functions
 is not the same as above (it depends on the
 dimension of $S_f$).

 Finally, the question of integrability must be
 considered, and similar considerations as above must
 be made for the submanifold $S_f$ instead of $J^{1*}E$.
 \item
 ({\sl Hamiltonian system associated with a hyper-regular Lagrangian system\/})

 If the Hamiltonian system $\hs$ is associated with a
 {\sl hyper-regular Lagrangian system}, then there exists the so-called
 {\sl Legendre map}, which is a diffeomorphism between $J^1E$ and $J^{1*}E$
 \cite{3}, \cite{8}, \cite{16}. In this case,
 it can be proved \cite{8}
 that, if $X_{\Lag}\in\vf^m(J^1E)$ and $X_{\cal H}\in\vf^m(J^{1*}E)$ are
 multivector fields solution of the Lagrangian and Hamiltonian
 field equations respectively, then
 $$
 \Lambda^m\Tan F\Lag\circ X_{\Lag}=fX_{\cal H}\circ F\Lag
 $$
 for some $f\in\Cinfty(J^{1*}E)$.
 That is, we have the following (commutative) diagram:
   $$
\begin{array}{ccc}
\Lambda^m\Tan J^1E & $\rightarrowfill$ & \Lambda^m\Tan J^{1*}E
\\
& \Lambda^m\Tan F\Lag &
\\
X_{\Lag}\ \Big\uparrow & &\Big\uparrow\ X_{\cal H}
\\
& F\Lag &
\\
J^1E & $\rightarrowfill$ & J^{1*}E
\end{array}
$$
(we say that the classes $\{ X_{\Lag}\}$ and $\{ X_{\cal H}\}$
 are $F\Lag$-related).
 \eit

 \section{Conclusions and outlook}

 We have used multivector fields in fiber bundles
 for setting and studying the Lagrangian and Hamiltonian field equations
 of first-order classical field theories. In particular, we have showed that:
\begin{itemize}
\item
The field equations for first order
 classical field theories
in the Lagrangian formalism ({\sl Euler-Lagrange equations}\/)
can be written using multivector fields in $J^1E$.
This description allow us to write the field equations for field
theories in an analogous way to the dynamical equations for
(time-dependent) Lagrangian mechanical systems.
\item
The Lagrangian equations can have no integrable solutions in $J^1E$,
for neither regular nor singular Lagrangian systems.

In the regular case, {\sl Euler-Lagrange multivector fields}
(that is, semiholonomic and solution of the equation
$\inn(X_{\Lag})\Omega_\Lag=0$)
always exist; but they are not necessarily integrable.
In the singular case, not even the existence of
such an Euler-Lagrange multivector field is assured.
In both cases, the multivector field solution (if it exists) is not unique.
 \item
 The Hamiltonian field equations
 can be written using multivector fields in $J^{1*}E$
 (the multimomentum bundle of the Hamiltonian formalism)
 in an analogous way to the dynamical equations for
 (time-dependent) Hamiltonian mechanical systems.
 \item
 The field equations $\inn(X_{\cal H})\Omega_h=0$,
 with $X_{\cal H}\in\vf^m(J^{1*}E)$ locally decomposable
 and $\bar\tau^1$-transverse,
 have solution everywhere in $J^{1*}E$, which is not
 unique; that is, there are classes of
 {\sl Hamilton-De Donder-Weyl multivector fields}
 which are solution of these equations. Nevertheless,
 these multivector fields
 are not necessarily integrable everywhere in $J^{1*}E$.
 \item
 This multivector field formulation is specially useful
 for characterizing {\sl symmetries}, both in the Lagrangian and
 Hamiltonian formalisms of field theories.
 First attempts in this subject have been already done
 \cite{7}, but new
 developments in this area are expected to be reached in the future.
 \eit

\subsection*{Acknowledgments}

We are grateful for the financial support of the CICYT PB98-0920.

\end{document}